\documentclass[12pt]{article}
\begin{document}

\begin{center}
{\Large\bf{}Gravitational energy in small regions for the
quasilocal expressions in orthonormal frames}
\end{center}

\begin{center}
Lau Loi So\\
Department of Physics, Tamkang University, Tamsui 251, Taiwan\\
(Dated on 23 September 2008)
\end{center}

\begin{abstract}
The M$\o$ller tetrad gravitational energy-momentum expression was
recently evaluated for a small vacuum region using orthonormal
frames adapted to Riemann normal coordinates.  However the result
was not proportional to the Bel-Robinson tensor
$B_{\alpha\beta\mu\nu}$. Treating a modified quasilocal
expressions in a similar way, we found one unique combination that
gives a multiple of $B_{\alpha\beta\mu\nu}$ which provides a
non-negative gravitational energy-momentum in the small sphere
approximation. Moreover, in addition to $B_{\alpha\beta\mu\nu}$,
we found a certain tensor
$S_{\alpha\beta\mu\nu}+K_{\alpha\beta\mu\nu}$ which gives the same
``energy-momentum" density in vacuum.  Using this tensor
combination, we obtained an infinite set of solutions that
provides a positive gravitational energy within the same limit.
\end{abstract}

\section{Introduction}
Finding an appropriate quasilocal expression for the gravitational
energy is an open problem in general relativity.  How do we even
know that gravitational energy exists?  Tidal friction from the
tidal force of the Moon on the Earth leads to the slowing of the
rate of Earth rotation.  The lengthing of the day is the indirect
evidence for the gravitational energy.

All matter and all other interaction fields, through their
energy-momentum density, act as the source of gravity. These
sources exchange energy-momentum with the gravitational field; all
attempts to identify an energy-momentum density for gravity itself
led to reference frame dependent quantities (i.e. pseudotensors),
a reflection of the fact that energy-momentum of an isolated
gravitating system is inherently non-local. This feature can be
understood in terms of the equivalence principle: gravity cannot
be detected at a point.  However, one can get around this
difficulty using the idea of quasilocal energy-momentum, i.e.,
associated with a closed 2-surface surrounding a region
\cite{Szabados}.

The Bel-Robinson tensor has a desirable property as it provides a
non-negative value, although it cannot be interpreted as a
``stress energy" for gravity directly because it has the wrong
dimension. The energy density has dimension cm$^{-2}$, while
$B_{\alpha\beta\mu\nu}t^{\alpha}t^{\beta}t^{\mu}t^{\nu}$ has
dimension cm$^{-4}$, where $t^{\alpha}$ is the timelike unit
normal.  However, the small sphere region limit can resolve this
mismatch.  In the small sphere limit, the quasilocal expression
for the energy-momentum density should be a multiple of
$r^{5}B_{\mu{}000}\sim\frac{4}{3}\pi{}r^{3} (r^{2}B_{\mu{}000})$
\cite{Szabados}, where $r$ is the radius of the Euclidean volume
3-ball. Indeed $r^{2}B_{0000}$ matches the energy density
dimension.

A positivity of gravitational energy proof was obtained in
orthonormal frames \cite{Nester1}. Success on a large scale
automatically implies the positivity on the small region limit.
Recently M$\o$ller's tetrad gravitational energy-momentum
expression \cite{Nester} was evaluated for a small vacuum region
using orthonormal frames adapted to Riemann normal coordinates.
This result for the gravitational energy in the small sphere
approximation is not positive definite. Treating a modified
quasilocal expression \cite{So} in a similar way, we found one
unique combination that gives a multiple of the Bel-Robinson
tensor which means that the gravitational energy is definitely
non-negative.

The Bel-Robinson tensor component $B_{0000}$ or $B_{00l}{}^{l}$
gives the non-negative gravitational ``energy" density.  We
discovered that the sum of the tensor components
$S_{0000}+K_{0000}$ or $S_{00l}{}^{l}+K_{00l}{}^{l}$ offers the
same ``energy" density value. Based on this criteria, we also
obtained an infinite set of solutions that provide a positive
gravitational energy within the same small sphere limit using
certain modified quasilocal expressions.

\section{Orthonormal frames and quadratic curvature}
The quasilocal quantities for small regions can be studied by
Taylor expanding the Hamiltonian, including the divergence of its
boundary term in a small spatial region surrounding a point. The
reference is the flat space geometry at this origin. The
orthonormal frame satisfies
\begin{eqnarray}
e^{\alpha}{}_{a}(0)=\delta^{\alpha}_{a},\quad\quad\quad~~{}
\partial_{i}e^{\alpha}{}_{a}(0)=0,\quad\quad\quad~~{}
\Gamma^{\alpha}{}_{\beta{}i}(0)=0,\\
\partial^{2}_{ij}e^{\alpha}{}_{a}(0)
=-\frac{1}{6}(R^{\alpha}{}_{iaj}+R^{\alpha}{}_{jai}),\quad~~
\partial_{j}\Gamma^{\alpha}{}_{\beta{}i}(0)
=\frac{1}{2}R^{\alpha}{}_{\beta{}ji},
\end{eqnarray}
where the Latin letters refer to coordinate frames (holonomic
frames) and Greek letter means the orthonormal frames
(non-holonomic frames).

The Bel-Robinson \cite{MTW} tensor is defined as
\begin{eqnarray}
B_{\alpha\beta\mu\nu}&:=&R_{\alpha\lambda\mu\sigma}R_{\beta}{}^{\lambda}{}_{\nu}{}^{\sigma}
+\ast{}R_{\alpha\lambda\mu\sigma}\ast{}R_{\beta}{}^{\lambda}{}_{\nu}{}^{\sigma}\nonumber\\
&=&R_{\alpha\lambda\mu\sigma}R_{\beta}{}^{\lambda}{}_{\nu}{}^{\sigma}
+R_{\alpha\lambda\nu\sigma}R_{\beta}{}^{\lambda}{}_{\mu}{}^{\sigma}
-\frac{1}{2}g_{\alpha\beta}R_{\lambda\sigma\rho\mu}R^{\lambda\sigma\rho}{}_{\nu},
\end{eqnarray}
where the dual curvature is $\ast{}R_{\alpha\beta\mu\nu}
=\frac{1}{2}\epsilon_{\alpha\beta\lambda\sigma}R^{\lambda\sigma}{}_{\mu\nu}$.
Furthermore, the tensors $S_{\alpha\beta\mu\nu}$ \cite{MTW} and
$K_{\alpha\beta\mu\nu}$ in vacuum are defined as
\begin{eqnarray}
S_{\alpha\beta\mu\nu}
:=R_{\alpha\mu\lambda\sigma}R_{\beta\nu}{}{}^{\lambda\sigma}
+R_{\alpha\nu\lambda\sigma}R_{\beta\mu}{}{}^{\lambda\sigma}
+\frac{1}{4}g_{\alpha\beta}g_{\mu\nu}
R_{\lambda\sigma\rho\tau}R^{\lambda\sigma\rho\tau},\\
K_{\alpha\beta\mu\nu}
:=R_{\alpha\lambda\beta\sigma}R_{\mu}{}^{\lambda}{}_{\nu}{}^{\sigma}
+R_{\alpha\lambda\beta\sigma}R_{\nu}{}^{\lambda}{}_{\mu}{}^{\sigma}
-\frac{3}{8}g_{\alpha\beta}g_{\mu\nu}
R_{\lambda\sigma\rho\tau}R^{\lambda\sigma\rho\tau}.
\end{eqnarray}
The identity in vacuum
$R_{\lambda\sigma\rho\mu}R^{\lambda\sigma\rho}{}_{\nu}
=\frac{1}{4}g_{\mu\nu}R_{\lambda\sigma\rho\tau}R^{\lambda\sigma\rho\tau}$
is useful.

It is known that the Bel-Robinson tensor is completely symmetric,
we have found the identity
\begin{eqnarray}
3B_{\alpha\beta\mu\nu}&\equiv&B_{\alpha\beta\mu\nu}+B_{\alpha\mu\beta\nu}+B_{\alpha\nu\beta\mu}\nonumber\\
&\equiv&S_{\alpha\beta\mu\nu}+S_{\alpha\mu\beta\nu}+S_{\alpha\nu\beta\mu}
+K_{\alpha\beta\mu\nu}+K_{\alpha\mu\beta\nu}+K_{\alpha\nu\beta\mu}.
\end{eqnarray}
Technically, the alternative form can be written as
\begin{equation}
B_{\alpha\beta\mu\nu}\equiv{}B_{\alpha(\beta\mu\nu)}\equiv{}S_{\alpha(\beta\mu\nu)}+K_{\alpha(\beta\mu\nu)}.
\end{equation}
The tensors $S_{\alpha\beta\mu\nu}$ and $K_{\alpha\beta\mu\nu}$
are both symmetric at the first pair and last pair of indices.

It turns out that in vacuum the small region energy density has a
RNC Taylor series expansion of the form
\begin{equation}
t^{\mu}{}_{\nu}=t^{\mu}{}_{\nu{}ij}x^{i}x^{j},
\end{equation}
the corresponding energy-momentum is
\begin{eqnarray}
P_{\mu}&=&\int_{t=0}t^{\nu}{}_{\mu{}ij}x^{i}x^{j}d\Sigma_{\nu}\nonumber\\
&=&t^{0}{}_{\mu{}ab}\int_{t=0}x^{a}x^{b}d^{3}x\nonumber\\
&=&t^{0}{}_{\mu{}ab}\frac{\delta^{ab}}{3}\int{}r^{2}d^{3}x\nonumber\\
&=&t^{0}{}_{\mu{}a}{}^{a}\,\frac{4\pi{}r^{5}}{15},
\end{eqnarray}
where $a,b=1,2,3$.  Note that
$t^{0}{}_{\mu{}a}{}^{a}=t^{0}{}_{\mu\alpha}{}^{\alpha}-t^{0}{}_{\mu{}0}{}^{0}$.
In particular $B^{0}{}_{ \mu{}a}{}^{a}=B^{0}{}_{ \mu{}00}$ as
$B_{\alpha\beta\mu\nu}$ is completely traceless.  More covariantly
\begin{equation}
t_{0\mu{}00}:=t_{\alpha\mu\beta\gamma}t^{\alpha}t^{\beta}t^{\gamma}.
\end{equation}
The ``energy-momentum" associated with the Bel-Robinson tensor is
\begin{eqnarray}
B_{\alpha\beta\mu\nu}t^{\beta}t^{\mu}t^{\nu}
=(E_{ab}E^{ab}+H_{ab}H^{ab},2\epsilon_{c}{}^{ab}E_{ad}H^{d}{}_{b}),
\end{eqnarray}
where the electric part $E_{ab}$ and magnetic part $H_{ab}$ are
defined in terms of the Weyl tensor as follows:
\begin{equation}
E_{ab}:=C_{ambn}t^{m}t^{n},\quad{}H_{ab}:=*C_{ambn}t^{m}t^{n}.
\end{equation}

Moreover for $S_{\alpha\beta\mu\nu}+K_{\alpha\beta\mu\nu}$, we
have the following identity related with the Bel-Robinson tensor
components
\begin{equation}
S_{\mu{}000}+K_{\mu{}000}\equiv{}B_{\mu{}000}\equiv{}B_{\mu{}0l}{}^{l}
\equiv{}S_{\mu{}0l}{}^{l}+K_{\mu{}0l}{}^{l}.\label{20May2008}
\end{equation}
It means that $S_{\mu{}000}+K_{\mu{}000}$ or
$S_{\mu{}0l}{}^{l}+K_{\mu{}0l}{}^{l}$ have the same physical
quantities as $B_{\mu{}000}$ or $B_{\mu{}0l}{}^{l}$.

\section{Modified quasilocal boundary expressions}
For a first order Lagrangian density:
\begin{equation}
{\cal{}L}=dq\wedge{}p-\Lambda(q,p)
\end{equation}
where $q$ is an f-form, $p$ is a 3-form and $\Lambda$ is the
potential.  The modified quasilocal expressions
\cite{So,Chen-Nester} can be briefly summarized as follows
\begin{equation}
{\cal{}B}_{c_{1},c_{2}}(N)={\cal{}B}_{p}(N)
+c_{1}i_{N}\Delta{}q\wedge\Delta{}p
+\epsilon{}c_{2}\Delta{}q\wedge{}i_{N}\Delta{}p,\label{29Jan2008}
\end{equation}
where
${\cal{}B}_{p}(N)=i_{N}\overline{q}\wedge\Delta{}p-\epsilon\Delta{}q\wedge{}i_{N}p$,
$\epsilon=(-1)^{f}$ with $f$-form, $c_{1}$ and $c_{2}$ are real
numbers, $\Delta{}q=q-\overline{q}$, $\Delta{}p=p-\overline{p}$,
$\overline{q}$ and $\overline{p}$ are the background reference
values. For GR
\begin{equation}
{\cal{}L}=R^{\alpha}{}_{\beta}\wedge\eta_{\alpha}{}^{\beta},
\end{equation}
so let
\begin{equation}
q\rightarrow\Gamma^{\alpha}{}_{\beta},\quad{}
p\rightarrow\frac{1}{2\kappa}\eta_{\alpha}{}^{\beta}.
\end{equation}
Allowing for the background connection
$\overline{\Gamma}^{\alpha}{}_{\beta}=0$, then (\ref{29Jan2008})
becomes
\begin{equation}
2\kappa\,{\cal{}B}_{c_{1},c_{2}}(N)=\Gamma^{\alpha}{}_{\beta}\wedge{}i_{N}\eta_{\alpha}{}^{\beta}
+c_{1}i_{N}\Gamma^{\alpha}{}_{\beta}\wedge\Delta\eta_{\alpha}{}^{\beta}
-c_{2}\Gamma^{\alpha}{}_{\beta}\wedge{}i_{N}\Delta\eta_{\alpha}{}^{\beta}.\label{25aMarch2008}
\end{equation}
When $(c_{1},c_{2})=(0,0)$, $(0,1)$, $(1,0)$ and $(1,1)$, the four
quasilocal boundary expressions with the simplest boundary
conditions are
\begin{eqnarray}
{\cal{}B}_{p}(0,0)&=&\Gamma^{\alpha}{}_{\beta}\wedge{}i_{N}\eta_{\alpha}{}^{\beta},\\
{\cal{}B}_{\rm{}d}(0,1)&=&\Gamma^{\alpha}{}_{\beta}\wedge i_{N}\overline{\eta}_{\alpha}{}^{\beta}, \\
{\cal{}B}_{\rm{}c}(1,0)&=&\Gamma^{\alpha}{}_{\beta}\wedge{}i_{N}\eta_{\alpha}{}^{\beta}
+i_{N}\Gamma^{\alpha}{}_{\beta}\wedge\Delta\eta_{\alpha}{}^{\beta},\label{19dMay2008}\\
{\cal{}B}_{q}(1,1)&=&\Gamma^{\alpha}{}_{\beta}\wedge{}i_{N}\overline{\eta}_{\alpha}{}^{\beta}
+i_{N}\Gamma^{\alpha}{}_{\beta}\wedge\Delta\eta_{\alpha}{}^{\beta}.
\end{eqnarray}

In terms of the superpotential, rewrite the modified quasilocal
expressions (\ref{25aMarch2008}) as
\begin{equation}
2\kappa\,{\cal{}B}_{c_{1},c_{2}}(N)=-\frac{1}{2}N^{\mu}\left\{
U_{\mu}{}^{[ij]}+{}c_{1}U_{\mu}{}^{[ij]}-{}c_{2}U_{\mu}{}^{[ij]}
\right\}\epsilon_{ij}.\label{28Jan2008}
\end{equation}
The tetrad teleparallel gauge current expression in orthonormal
frames is
\begin{equation}
U_{\mu}{}^{[ij]}=-e\overline{g}^{\beta\sigma}
\Gamma^{\alpha}{}_{\beta{}m}\delta^{\rho\tau\gamma}_{\alpha\sigma\mu}
e^{m}{}_{\gamma}e^{i}{}_{\tau}e^{j}{}_{\rho},
\end{equation}
and in RNC
\begin{eqnarray}
c_{1}U_{\mu}{}^{[ij]}&=&-\frac{c_{1}}{12}e\overline{g}^{\beta\sigma}
R^{\alpha}{}_{\beta\gamma\mu}R^{\tau}{}_{\xi\kappa\lambda}
e^{i}{}_{\rho}e^{j}{}_{\pi}x^{\gamma}x^{\xi}x^{\kappa}\delta^{\rho\pi\lambda}_{\alpha\sigma\tau}
+{\cal{}O}(x^{4}),\\
c_{2}U_{\mu}{}^{[ij]}&=&
\frac{c_{2}}{12}e\overline{g}^{\beta\sigma}R^{\alpha}{}_{\beta\gamma\lambda}R^{\nu}{}_{\xi\kappa\tau}
e^{i}{}_{\rho}e^{j}{}_{\pi}x^{\gamma}x^{\xi}x^{\kappa}\delta^{\rho\pi\lambda\tau}_{\alpha\sigma\mu\nu}
+{\cal{}O}(x^{4}).
\end{eqnarray}
It should be noted that both the tetrad teleparallel gauge current
$U_{\mu}{}^{[ij]}$ and the associated energy-momentum density
$\partial_{j}(U_{\mu}{}^{[ij]})$ is a tensor. In contrast, the
M{\o}ller 1961 expression $_{M}U_{h}{}^{[ij]}$ is a tensor but the
corresponding energy-momentum density
$\partial_{j}(_{M}U_{h}{}^{[ij]})$ is not a tensor.  Precisely it
is a pseudotensor, which means it depends on the coordinates in a
non-covariant way. As
$_{M}U_{h}{}^{[ij]}=e^{\mu}{}_{h}U_{\mu}{}^{[ij]}$, modify the
superpotential of (\ref{28Jan2008}) to
\begin{equation}
2\kappa\,{\cal{}U}_{h}{}^{[ij]}={}_{M}U_{h}{}^{[ij]}
+e^{\mu}{}_{h}\left({}c_{1}U_{
\mu}{}^{[ij]}-{}c_{2}U_{\mu}{}^{[ij]}\right).
\end{equation}
Then the corresponding pseudotensor becomes locally in a small
region in RNC
\begin{eqnarray}
2\kappa\,t_{h}{}^{i}&=&2\kappa\,\partial_{j}{\cal{}U}_{h}{}^{[ij]}\nonumber\\
&=&e\,2G_{h}{}^{i}\nonumber\\
&&+\frac{e}{24}\left\{ 2(2-3c_{2}) B_{h}{}^{i}{}_{\xi\kappa}
-(1-3c_{1}+3c_{2})S_{h}{}^{i}{}_{\xi\kappa}
+2(c_{1}-2c_{2})K_{h}{}^{i}{}_{\xi\kappa}
\right\}x^{\xi}x^{\kappa}\nonumber\\
&&+{\cal{}O}({\rm{}Ricci},x)+{\cal{}O}(x^{3}).\label{22May2008}\quad\quad
\end{eqnarray}
Regarding whether the above expression is good for the
gravitational energy, there are three limits we can consider. They
are inside matter (interior mass density), at spatial infinity
(ADM mass energy) and in vacuum (positive gravitational energy).
The first two tests are relatively mild in general, but the last
one is not, because it is very sensitive if we insist to obtain
the Bel-Robinson tensor. For example, the M{\o}ller 1961
expression fulfills
the first two tests while it fails the third examination \cite{Nester}.\\
Test (i): Inside matter. The energy density inside matter at the
origin is
\begin{equation}
{\cal{}E}=-t_{0}{}^{0}(0)=-\frac{G_{0}{}^{0}(0)}{\kappa}=-T_{0}{}^{0}(0)=\rho.
\end{equation}
Test (ii): At spatial infinity.  The total energy of the
pseudotensor agrees with the ADM mass formula \cite{ADM}
\begin{equation}
E=\frac{1}{2\kappa}\oint{}N^{0}U_{0}{}^{[\mu\nu]}\epsilon_{\mu\nu}
=\frac{1}{2\kappa}\lim_{r\rightarrow\infty}
\sum_{i,j=1}^{3}\oint\left(h_{ij,i}-h^{i}{}_{i,j}\right)N^{j}dA,
\end{equation}
where the integrals are taken over a sphere of constant $r$ and
$N^{j}=x^{j}/r$ is the outward normal to this sphere.\\
Test (iii): In vacuum.  Consider (\ref{22May2008}) by eliminating
the tensors $S_{\alpha\beta\mu\nu}$ and $K_{\alpha\beta\mu\nu}$
when $(c_{1},c_{2})=(\frac{2}{3},\frac{1}{3})$, the gravitational
energy-momentum density in the small sphere region limit is
\begin{equation}
t_{\alpha}{}^{\beta}
=\frac{1}{12\kappa}B_{\alpha}{}^{\beta}{}_{\xi\kappa}x^{\xi}x^{\kappa}.
\end{equation}
This is the first desired result we found in vacuum.  It is
proportional to the Bel-Robinson tensor which is an invariant
strength measurement of the non-negative gravitational energy
density within a very small region. Explicitly
\begin{equation}
B_{0000}=E_{ab}E^{ab}+H_{ab}H^{ab}\geq{}0.
\end{equation}

The second desired result for the non-negative gravitational
energy-momentum in the small sphere limit is
\begin{eqnarray}
P_{\mu}&=&(-E,\vec{P})\nonumber\\
&=&-\frac{1}{48\kappa}\int\left\{2(2-3c_{2})B_{\mu{}0ij}
-(1-3c_{1}+3c_{2})S_{\mu{}0ij} +2(c_{1}-2c_{2})K_{\mu{}0ij}
\right\}x^{i}x^{j}d^{3}x\nonumber\\
&=&-\frac{2c_{1}-1}{240G}\,r^{5}B_{\mu{}000},
\end{eqnarray}
provided $c_{1}\geq{}1/2$ and the unique combination
$c_{1}+c_{2}=1$, which is the constraint such that the
coefficients of $S_{\mu{}0l}{}^{l}$ and $K_{\mu{}0l}{}^{l}$ are
the same. There is an infinite set of solutions because of the
constant $c_{1}$. When $(c_{1},c_{2})=(1,0)$, the quasilocal
expression for this set is $B_{c}(1,0)$ as mentioned in
(\ref{19dMay2008}) which has a simple boundary condition.

\section{Conclusion}
Once the positivity energy proof is achieved for some particular
expression, the small sphere limit is guaranteed for the positive
gravitational energy calculation.  Recently, the M$\o$ller tetrad
gravitational energy-momentum expression was evaluated for a small
vacuum region using orthonormal frames adapted to Riemann normal
coordinates \cite{Nester}. Treating a modified quasilocal
expressions in a similar way, we found one unique combination that
gives a multiple of the Bel-Robinson tensor.

The components of the Bel-Robinson tensor $B_{0000}=B_{00l}{}^{l}$
gives the non-negative ``energy" density which has a nice property
for the gravitational field. However, besides this tensor, we
found that $S_{0000}+K_{0000}=S_{00l}{}^{l}+K_{00l}{}^{l}$ gives
the same physical value. Based on this property, we also obtained
an infinite set of solutions that provide the positive
gravitational energy within the same region limit using certain
modified quasilocal expressions.

\section*{Acknowledgment}
This work was supported by NSC 96-2811-M-032-001.



\begin{thebibliography}{3}

\bibitem{Szabados}
L.B. Szabados 2004 {\it{}Living Rev. Rel.} {\bf 7} 4

\bibitem{Nester1}
J.M. Nester 1989 {\it Phy. Lett. A} {\bf 139} 112

\bibitem{Nester}
L.L. So and J.M. Nester 2006 {\it{}Preprint} gr-qc/0612061

\bibitem{So}
L.L. So, {\it Int. J. Mod. Phys. D} 2007 $\mathbf{16}$  875

\bibitem{MTW}
C.W. Misner, K.S. Thorne and J.A. Wheeler 1973 {\it Gravitation}
(San Francisco: Freeman)

\bibitem{Chen-Nester}
C.M. Chen  and  J.M. Nester 1999 {\it Class. Quantum Grav.}  {\bf
16} 1279

\bibitem{ADM}
R. Arnowitt, S. Deser and C.W. Misner 1961 {\it{}Phy. Rev.}
{\bf{}122} 997

\end{thebibliography}
\end{document}